\documentclass[11pt,a4paper]{article} 
\pdfoutput=1
\usepackage{jcappub}
\usepackage{subfigure}
\newcommand{\eq}[1]{\begin{equation}#1\end{equation}}
\newcommand{\eqa}[1]{\begin{eqnarray}#1\end{eqnarray}}
\newcommand{\secs}[1]{\section{#1\label{sec-#1}}}
\newcommand{\ssecs}[1]{\subsection{#1\label{ssec-#1}}}

\newcommand{\fig}[4]{\begin{figure}[#4]\centering\includegraphics[width=#3\textwidth]{Graph-#1.pdf}\caption{#2}\label{fig-#1}\end{figure}}
\newcommand{\figa}[3]{\begin{figure}[#3]\centering #1\caption{#2}\end{figure}}
\newcommand{\figi}[3]{\subfigure[#2]{\includegraphics[width=#3\textwidth]{Graph-#1.pdf}\label{fig-#1}}}

\newcommand{\refeq}[1]{eq.\ (\ref{eq-#1})}
\newcommand{\refsec}[1]{section \ref{sec-#1}}
\newcommand{\refssec}[1]{section \ref{ssec-#1}}

\newcommand{\refig}[1]{figure \ref{fig-#1}}

\newcommand{\Refeq}[1]{Eq.\ (\ref{eq-#1})}

\newcommand{\subs}[1]{_\mathrm{#1}}
\newcommand{\sups}[1]{^\mathrm{#1}}
\newcommand{\dd}[1]{\mathrm{d}#1}

\newcommand{\cm}[1]{}
\newcommand{\Gaunt}[6]{G_{#1}^{#4}{}_{#2}^{#5}{}_{#3}^{#6}}
\newcommand{\Ga}[3]{\Gaunt{l_{#1}}{l_{#2}}{l_{#3}}{m_{#1}}{m_{#2}}{m_{#3}}}
\newcommand{\Wigner}[6]{\left(\begin{array}{ccc}#1&#2&#3\\#4&#5&#6\end{array}\right)}
\newcommand{\Sj}[6]{\left\{\begin{array}{ccc}#1&#2&#3\\#4&#5&#6\end{array}\right\}}

\newcommand{\Wj}[3]{\Wigner{l_{#1}}{l_{#2}}{l_{#3}}{m_{#1}}{m_{#2}}{m_{#3}}}
\newcommand{\Wjz}[3]{\Wigner{l_{#1}}{l_{#2}}{l_{#3}}{0}{0}{0}}
\newcommand{\vect}[1]{\mathbf{#1}}
\newcommand{\uv}[1]{\vect{\hat{#1}}}

\def\cc{(cite \#) }
\def\fNL{f\subs{NL}}
\def\tNL{\tau\subs{NL}}
\def\gNL{g\subs{NL}}

\begin{document}
\title{Detecting CMB modulations and non-Gaussianities from power multipoles}
\author{Lingfei Wang}
\affiliation{Physics Department, Lancaster University, Lancaster LA1 4YB, UK}
\abstract{By decomposing the CMB temperature fluctuations \emph{squared} into spherical harmonics, we can define the \emph{CMB power multipoles} as the fundamental observables to quantify any generic modulations in the CMB. This allows for simple direct measurements for the CMB modulations, and also indirect measurements for the CMB trispectrum, which fully determines the CMB modulations in a statistically isotropic universe. The CMB power multipoles can reach the theoretical accuracies higher than the current constraints from Planck observations. The theoretical accuracy for the dipole modulation, $A\sim0.0015$, receives $\sim 10$ times improvement; higher $l$ modulations have even higher accuracies. This calls for practical observational studies of its possible improvements on the CMB modulation and non-Gaussianity constraints.}
\maketitle
\secs{Introduction}
The cosmic microwave background (CMB) is the most precise observation we can currently perform on the earliest ages of our universe. Following the WMAP satellite and its successor Planck, we are entering the stage of ``precision cosmology'', so the history of our universe, especially inflation, can be highly constrained. The Planck observations have detected no evidence yet of the primordial non-Gaussianities \cite{Ade:2013ydc} or isocurvature perturbations \cite{Ade:2013uln}, while confirming the small spectral tilt in the primordial curvature perturbations \cite{Ade:2013zuv}. We have once again verified the standard paradigm of cosmology.

These progresses are achieved by analyzing the angular multipoles of the CMB temperature fluctuations. Assuming statistical isotropy, we can extract the statistical information from its angular correlation functions, such as angular power spectrum, bispectrum, and trispectrum \cite{Gangui:1993tt,Hu:2001fa}. We have been able to learn the $C_l$ curve for the angular spectrum, but for the CMB bispectrum and trispectrum, we have only been able to measure some specific shapes of the angular correlation functions, such as the local, equilateral and orthogonal $\fNL$, and $\tNL$. Much of the information on the CMB bispectrum and trispectrum remains unknown, partly because of the huge computational cost of the data analysis of high-resolution observations. For example, there is still a wide range of possibilities that the universe deviates the standard paradigm in some other way, such as a large non-Gaussianity in nonconventional shapes \cite{Regan:2010cn}.

\cm{but most of the information on bispectrum and trispectrum still remains unknown. This is partly because a complete data analysis would require the computer time $\sim{\cal O}(l\subs{max}^5)$ for the bispectrum, and $\sim{\cal O}(l\subs{max}^7)$ for the trispectrum, which are unrealistic even on today's super computers for a high-resolution observation like Planck with $l\subs{max}\sim2500$. Till now, we have only learned about the CMB bispectrum $\fNL$ of the local, equilateral and orthogonal types, and the local trispectrum $\tNL$. There is still a wide range of possibilities that the universe deviates the standard paradigm in some other way, such as a large non-Gaussianity in nonconventional shapes\cc.}

Meanwhile, another possible deviation that has received much attention is the CMB power asymmetry or dipole modulation \cite{Ade:2013nlj}. Such a dipole modulation or even higher multipole modulations certainly act as tests for the standard paradigm of cosmology by themselves. On the other hand, theoretical investigations have found they are closely related to the primordial non-Gaussianities of the CMB \cite{Lyth:2013vha,Wang:2013lda,Namjoo:2013fka}.
\footnote{For previous studies, see \cite{Eriksen:2003db,Eriksen:2007pc,Erickcek:2008sm,Erickcek:2008jp,Hoftuft:2009rq,Hirata:2009ar}; for more recent ones, see \cite{Chang:2013vla,Dai:2013kfa,Liu:2013kea,McDonald:2013aca,Chen:2013eaa,Liddle:2013czu,Mazumdar:2013yta,Cai:2013gma,Flender:2013jja,Kohri:2013kqa,McDonald:2013qca,Kanno:2013ohv,Chang:2013mya}.}
This suggests the CMB modulations can also serve as indirect detections for primordial non-Gaussianities, although the current focus has been mostly on dipole modulations only.

A CMB modulation should give rise to the temperature fluctuations $\Delta T(\uv n)$ whose amplitude depends on the direction $\uv n$, while we can also measure its amplitude at any direction easily by averaging $\Delta T^2(\uv n)$ locally. In this sense, we can quantify the strength of the modulation with the \emph{power multipoles} of the CMB, which is the spherical harmonic expansion of $\Delta T^2(\uv n)$, like our conventional CMB multipoles from the spherical harmonic expansion of $\Delta T(\uv n)$. After that, the power spectrum of the power multipoles, namely $C^{(2,2)}_l$, i.e.\ the counterpart of $C_l$, characterizes easily the modulation strength of the corresponding angular number $l$. The CMB power multipoles also act as indirect measurements for the CMB non-Gaussianities, which have a theoretical accuracy higher than the Planck measurement, and which is more capable in distinguishing the non-Gaussianity shapes.

I will first briefly mention the CMB multipole calculations for its power spectrum and trispectrum in \refsec{CMB multipoles --- a brief overview}. The CMB power multipoles will be defined, and demonstrated to relate to the CMB trispectrum in \refsec{CMB power multipoles}. The paper then calculates the natural amount of power multipoles from a Gaussian universe in \refssec{From Gaussian perturbations}, and the statistical error from cosmic variance in \refssec{From cosmic variance}, as the baseline for comparisons with observations. Excessive power multipoles beyond the error tolerance will then indicate a deviation from the standard paradigm of cosmology, which may come from primordial non-Gaussianities as calculated in \refssec{From non-Gaussian perturbations}. Comparisons will be made in \refsec{Comparison with CMB modulation models} with the current techniques for the modulation analysis, such as the power asymmetry factor $A$,  or the Bipolar Spherical Harmonics (BipoSH). The article concludes in \refsec{Conclusion}.

\secs{CMB multipoles --- a brief overview}
This paper will only consider a full sky and neglect secondary CMB effects or noises. The CMB temperature fluctuations, $\Delta T(\uv{n})$, are decomposed into spherical harmonics, $Y_{lm}(\uv{n})$, in the form
\eq{\Delta T(\uv{n})=\sum_{lm}a_{lm}Y_{lm}(\uv{n}).}
The temperature fluctuations were seeded by $\Phi(\mathbf{k})$, the primordial curvature perturbations generated during inflation. Assuming statistical isotropy, it can be shown that the angular multipole modes $a_{lm}$ satisfy
\eq{a_{lm}=(-i)^l\int\frac{\dd^3\vect k}{2\pi^2}\,Y_{lm}^*(\uv{k})g_l(k)\Phi(\mathbf{k}),\label{eq-bo-alm}}
where $g_l(k)$ is the transfer function for the perturbations. Its angular power spectrum can then be calculated as
\eq{\langle a_{lm}^*a_{l'm'}\rangle=C_l\delta_{ll'}\delta_{mm'},\label{eq-bo-C}}
where
\eq{C_l=\frac{2}{\pi}\int\dd k\,k^2|g_l(k)|^2P_\Phi(k).}
Here we have used the power spectrum of primordial curvature perturbations $\langle\Phi(\mathbf{k})\Phi^*(\mathbf{k'})\rangle=8\pi^3\delta^3(\mathbf{k}-\mathbf{k'})P_\Phi(k)$.

We will also be interested in the angular trispectrum of the CMB multipoles, which can be separated into a non-connected part and a connected part, as
\eq{\langle a_{l_1m_1}a_{l_2m_2}a_{l_3m_3}a_{l_4m_4}\rangle=\langle a_{l_1m_1}a_{l_2m_2}a_{l_3m_3}a_{l_4m_4}\rangle\subs{nc}+\langle a_{l_1m_1}a_{l_2m_2}a_{l_3m_3}a_{l_4m_4}\rangle\subs c.\label{eq-bo-trisep}}
In a Gaussian universe, the connected part vanishes while the non-connected part gains a contribution
\eq{\langle a_{l_1m_1}a_{l_2m_2}a_{l_3m_3}a_{l_4m_4}\rangle\subs{nc}=(-1)^{m_1+m_3}C_{l_1}C_{l_3}\delta_{l_1l_2}\delta_{l_3l_4}\delta_{m_1-m_2}\delta_{m_3-m_4}+\underset{l_2\leftrightarrow l_3,l_4}{2\mathrm{\ perms}}.\label{eq-bo-trinc}}

Non-Gaussianities can be generated by primordial mechanisms or post-inflationary effects. In either case, they will contribute to the connected part as well as the non-connected part. In our near-Gaussian universe, we only consider tree-level contributions from the non-Gaussianities, in which case the non-connected angular trispectrum remains unchanged. In case of any primordial non-Gaussianities, the contribution to the connected angular trispectrum, $\langle a_{l_1m_1}a_{l_2m_2}a_{l_3m_3}a_{l_4m_4}\rangle\subs c$, is calculated in \refsec{Angular trispectra from primordial non-Gaussianities}.

\secs{CMB power multipoles}
\figa{\figi{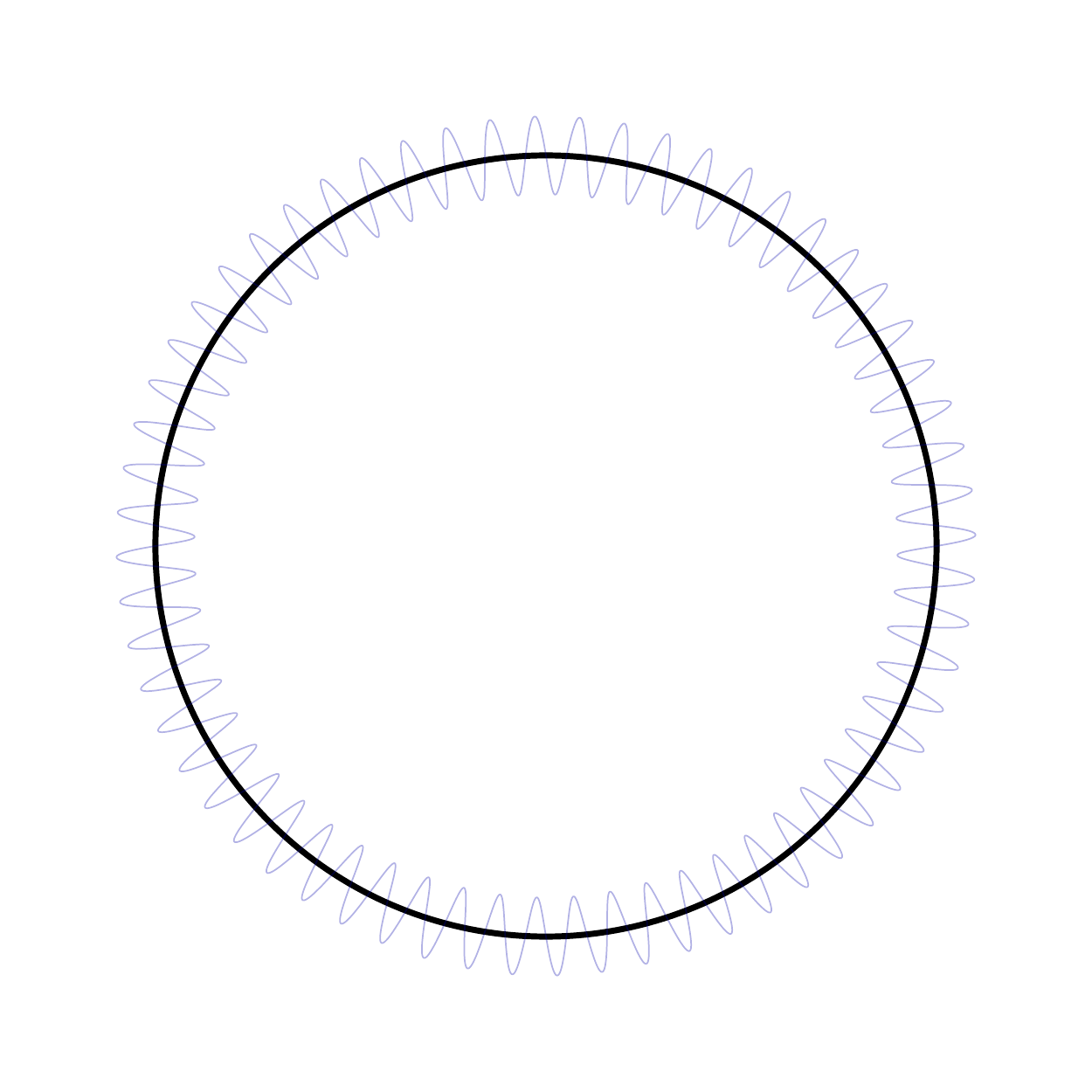}{No CMB modulation\newline in CMB space.}{0.33}\figi{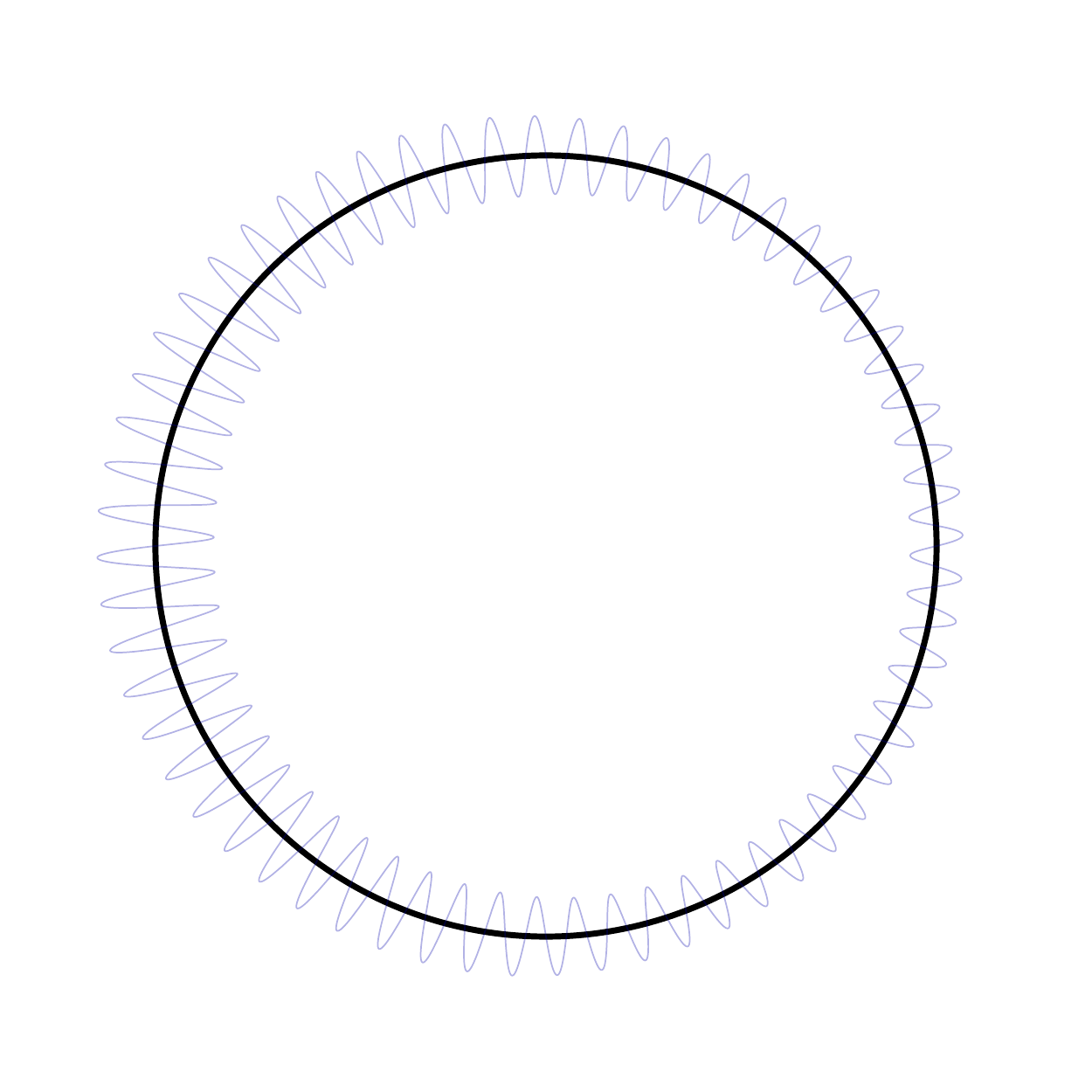}{Dipole CMB modulation\newline in CMB space.}{0.33}\figi{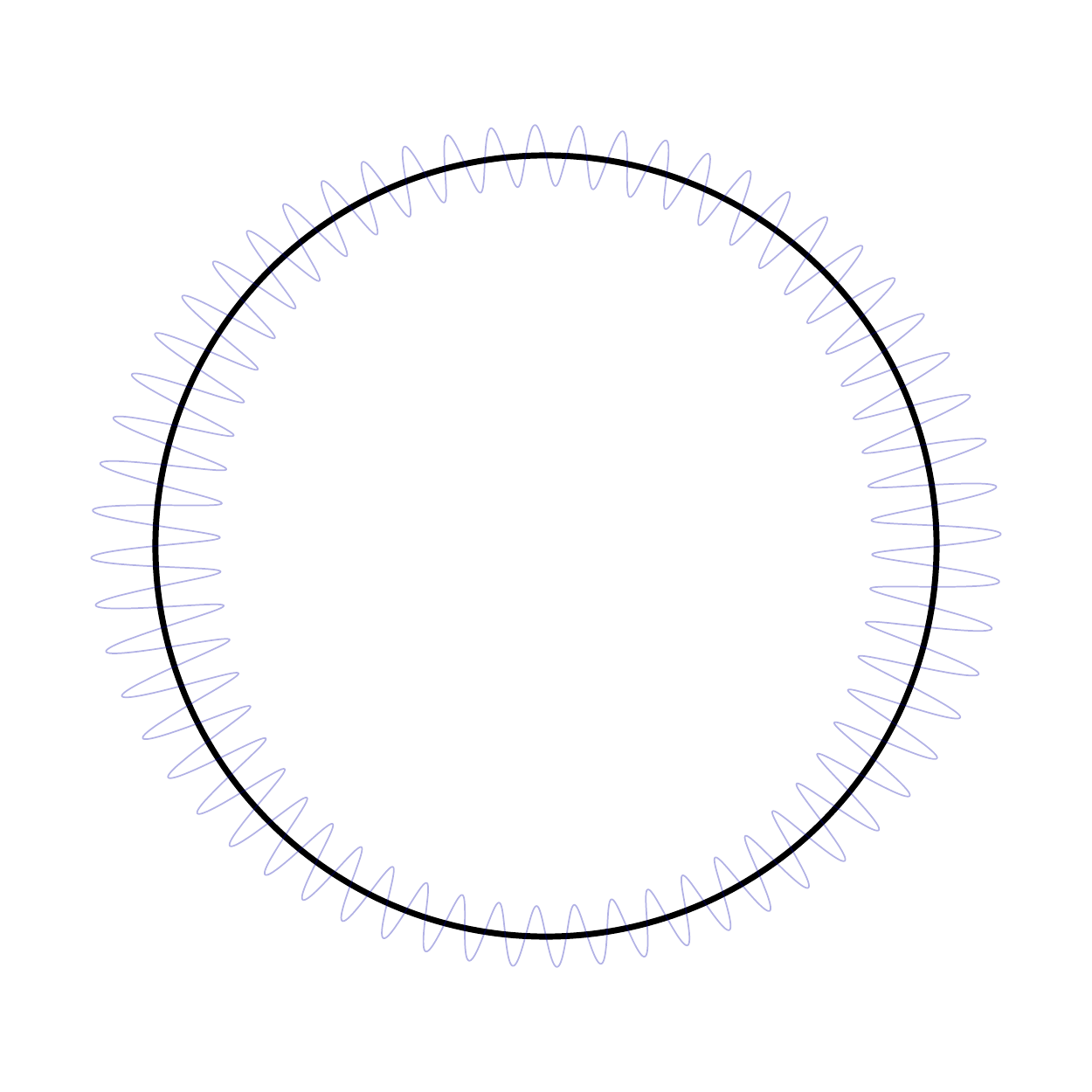}{Quadrupole CMB modulation\newline in CMB space.}{0.33}\\
\figi{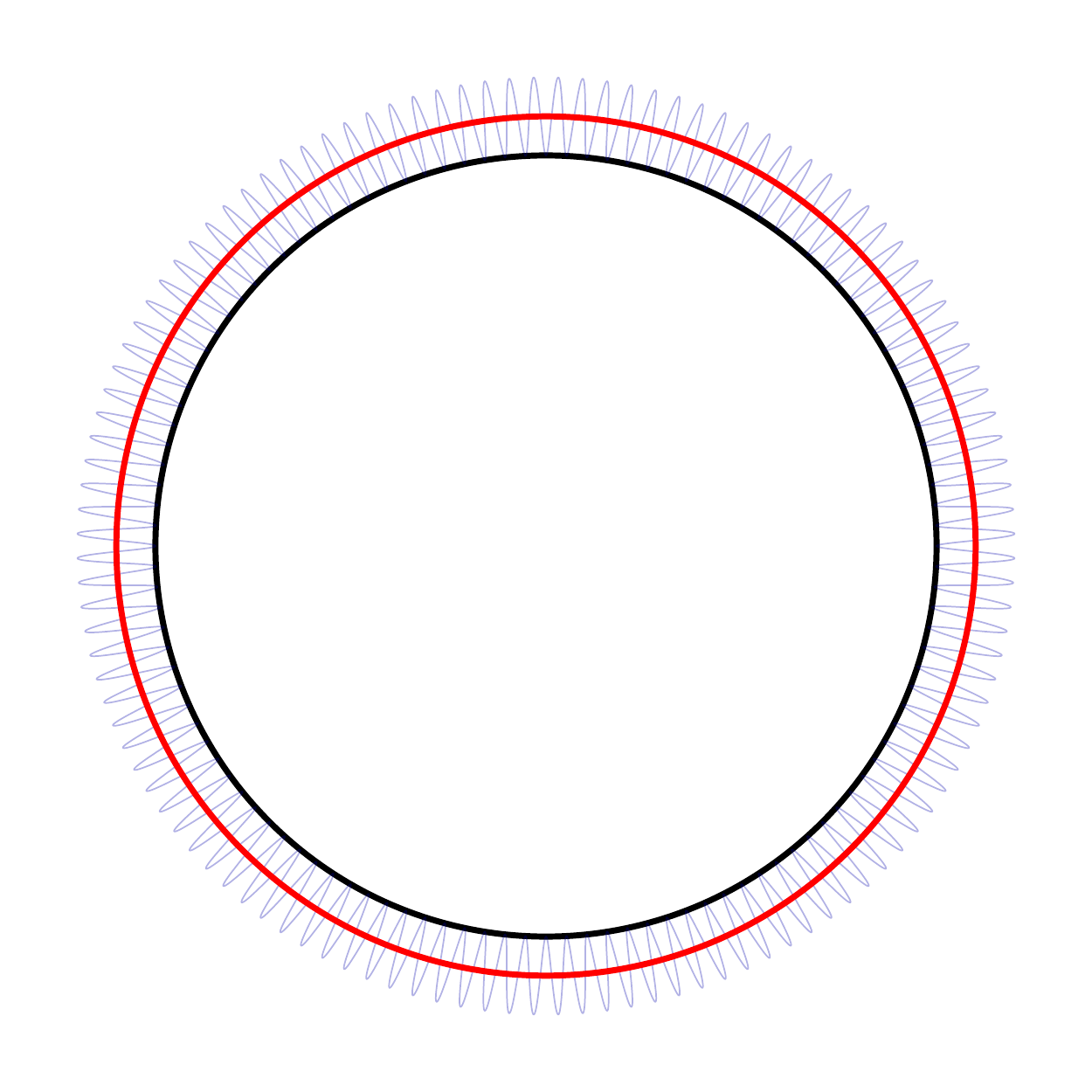}{No CMB modulation\newline in CMB power space.}{0.33}\figi{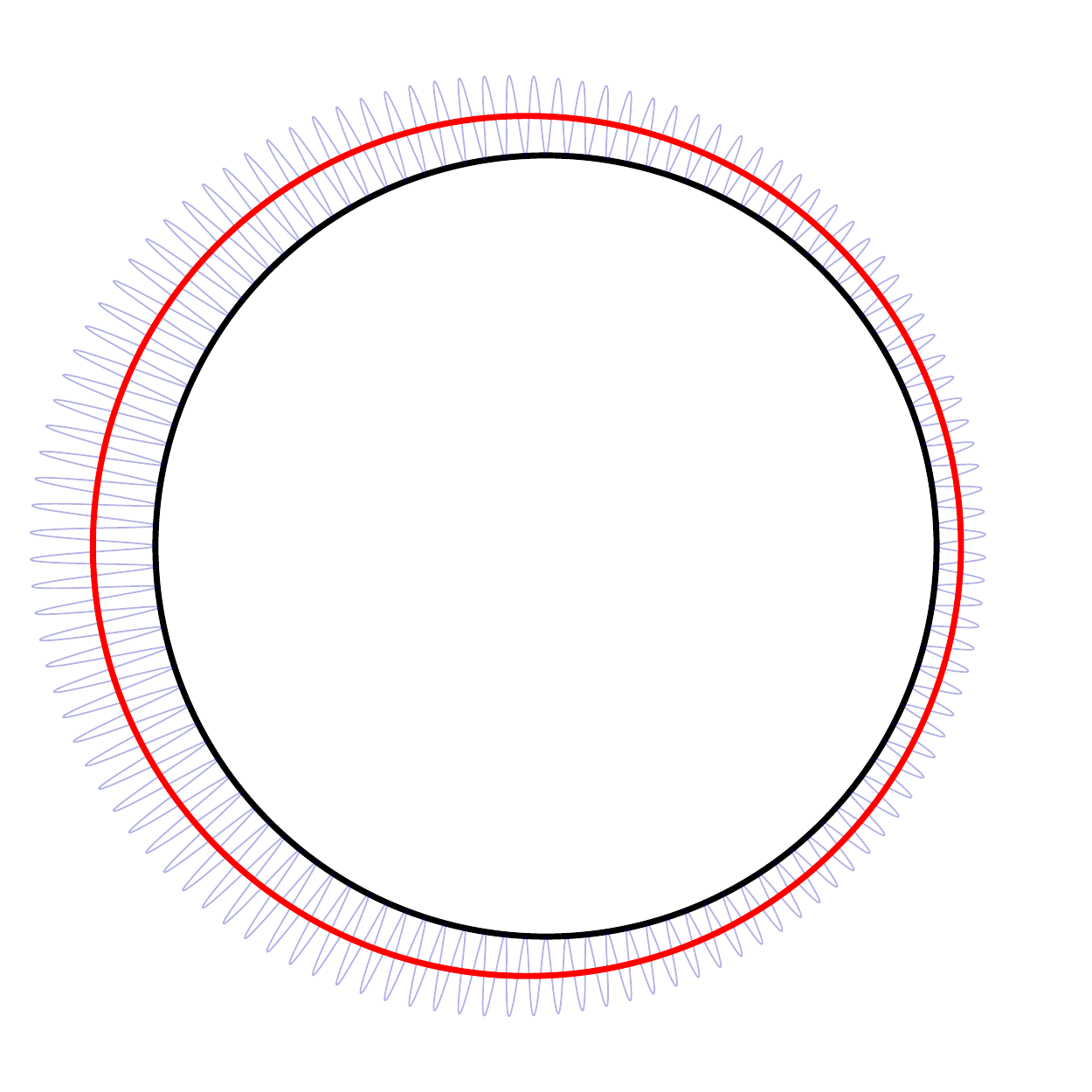}{Dipole CMB modulation\newline in CMB power space.}{0.33}\figi{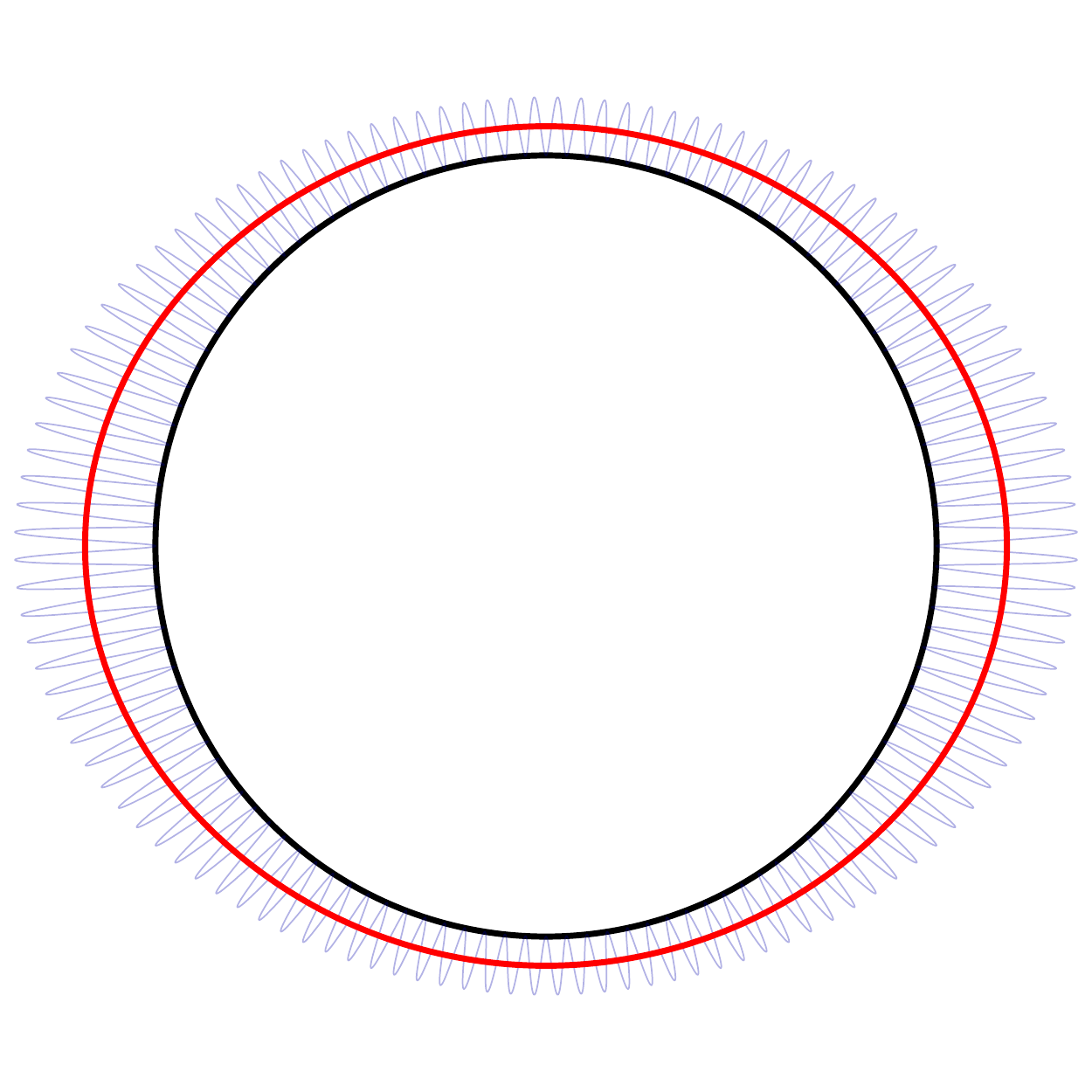}{Quadrupole CMB modulation\newline in CMB power space.}{0.33}}{A schematic demonstration of how modulations are manifested in the CMB space and the CMB power space. The black circles represent the CMB background temperature (power) and the blue curves are the CMB fluctuations or CMB fluctuation powers on top or at the bottom respectively. The red curves show the \emph{locally averaged} CMB powers, i.e. $\Delta T^2(\uv{n})$, which will pick up the corresponding angular multipoles when the CMB is modulated.\label{fig-comp}}{}
The CMB power multipoles are defined in a way such that it measures the modulations in the CMB perturbations. It can come from the natural fluctuations in the quantum system, or other effects such as nonlinear perturbations. These modulations can be measured from the power of the CMB temperature fluctuations or the \emph{CMB (fluctuation) power}, $\Delta T^2(\uv{n})$, whose local average represents the local power/amplitude of the perturbations, as shown in \refig{comp}. For example, in case of a significant power asymmetry in $\Delta T(\uv{n})$, we should see a corresponding dipole in $\Delta T^2(\uv{n})$. Similarly, modulations of higher multipoles in the CMB temperature fluctuations will also show up as the corresponding multipoles in the CMB power map for $\Delta T^2(\uv{n})$.

Following the formalism in \refsec{CMB multipoles --- a brief overview}, we can apply similar techniques to predict the angular multipoles of the CMB power. We define the power multipoles, $a_{lm}^{(2)}$, as
\eq{\Delta T^2(\uv{n})=\sum_{lm}a_{lm}^{(2)}Y_{lm}(\uv{n}).}
The CMB power multipoles are totally determined by the CMB multipoles following the relation\footnote{It should be noted that $a_{lm}^{(2)}$ are not Gaussian random variables. This however will not change any of the calculations.}
\eq{a_{lm}^{(2)}=(-1)^m\sum_{l_1l_2m_1m_2}a_{l_1m_1}a_{l_2m_2}\Gaunt{l_1}{l_2}{\ \ l}{m_1}{m_2}{-m},}
where the Gaunt integral is defined as
\eq{\Ga{1}{2}{3}\equiv\int\dd^2\uv{n}\,Y_{l_1m_1}(\uv{n})Y_{l_2m_2}(\uv{n})Y_{l_3m_3}(\uv{n}).}

Then we can compute the CMB modulation with the angular spectrum for $\Delta T^2(\uv{n})$
\eqa{\langle a_{lm}^{(2)*}a_{l'm'}^{(2)}\rangle=(-1)^{m'}\sum_{l_1l_2l_3l_4m_1m_2m_3m_4}\Gaunt{l_1}{l_2}{l}{m_1}{m_2}{m}\Gaunt{l_3}{l_4}{l'}{m_3}{m_4}{-m'}\langle a_{l_1m_1}a_{l_2m_2}a_{l_3m_3}a_{l_4m_4}\rangle.\label{eq-pm-pm}}
This paper assumes statistical isotropy in the universe ensemble, so any modulation should not have any \emph{prior} preferred direction. Then we can apply the rotational invariance of the observable $\Delta T^2(\uv{n})$, similarly to what we do for the temperature fluctuations $\Delta T(\uv{n})$ in \cite{Hu:2001fa}. The power multipoles $a_{lm}^{(2)}$ should also conform with the relation
\eq{\langle a_{lm}^{(2)*}a_{l'm'}^{(2)}\rangle=C^{(2,2)}_l\delta_{ll'}\delta_{mm'}.}
We can hence use the parameter $C^{(2,2)}_l$ to indicate the strength of the multipole modulations, such as $C^{(2,2)}_1$ would correspond to the dipole modulation as shown in \refig{comp}. Note that $C^{(2,2)}_l$ indicates the multipole modulation with the angular number $l$, which is different from the ``scale dependence'' of the CMB power asymmetry.

From \refeq{pm-pm}, we can also see that the power multipoles are solely determined by the angular trispectrum. They do not extract any \emph{extra} information beyond the CMB angular correlation functions, but they linearly recombine the angular trispectra for the sake of modulation detection. \Refeq{pm-pm} also confirms the previous analyses of the power asymmetries in \cite{Wang:2013lda}, although the bispectra only contribute indirectly to the power multipoles through the trispectra.

\ssecs{From Gaussian perturbations}
When the CMB temperature fluctuations are completely Gaussian, we can substitute the non-connected angular trispectrum \refeq{bo-trinc} into the power multipoles \refeq{pm-pm}. This will bring about
\eq{C\sups{(2,2,nc)}_l=\sum_{l_1l_2}\frac{(2l_1+1)(2l_2+1)}{4\pi}P_{l_1l_2l}C_{l_1}C_{l_2},\hspace{0.5in}\mathrm{for\ }l=1,2,3,\dots,\label{eq-pm-G-C}}
where ``nc'' stands for non-connected, and $C_{l}$ is the CMB angular spectrum defined in \refeq{bo-C}. Also,
\eq{P_{l_1l_2l}\equiv\int_{-1}^1P_{l_1}(x)P_{l_2}(x)P_l(x)\dd x,}
where $P_l(x)$ is the Legendre polynomial.

The dipole modulation in the CMB should be manifested in the power dipole, which has the angular spectrum
\eq{C\sups{(2,2,nc)}_1=\frac{1}{\pi}\sum_llC_lC_{l-1}.}
However, the ``power asymmetry'' should depend on all the $C\sups{(2,2,nc)}_l$ with odd $l$'s, which provide asymmetric modulations. The specific contributions from different $C\sups{(2,2,nc)}_l$ depend on the shape of the opposite patches chosen in order to measure the power asymmetry. The choice of relatively small patches typically allows larger contributions from high $l$ modulations, and therefore can significantly overestimate the power asymmetry. Choosing larger patches or even hemispheres, on the other hand, can localize the contributions at low $l$ better, although high $l$ contributions are still non-zero.

\fig{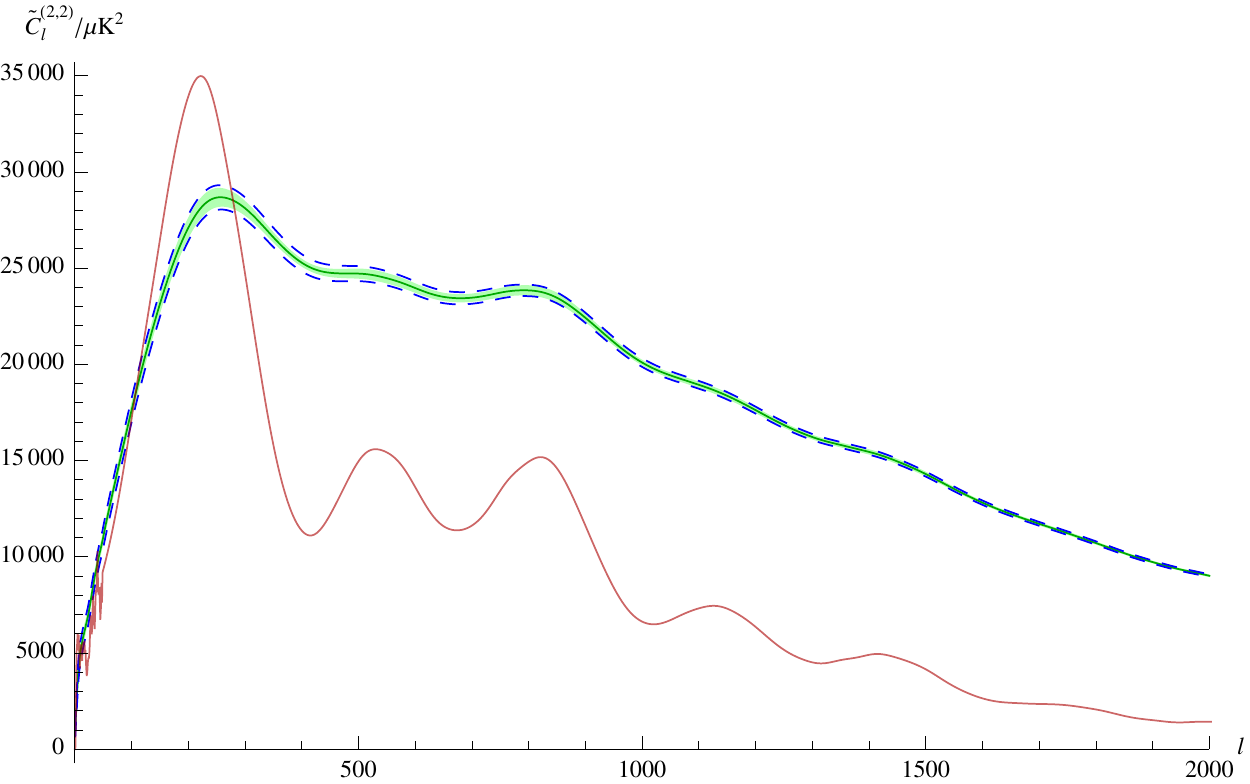}{The expectation of CMB power multipoles $\widetilde C\sups{(2,2)}_l$ (green solid curve) compared with the best-fit  of the CMB multipoles $\widetilde C_l$ (red solid curve), when assuming the $C_l$ curve of the universe ensemble is exactly the one we observe. The green shaded region is the standard deviation of cosmic variance, from the $10^5$ universe samples of the Monte Carlo simulation. The dashed blue curves are the estimated $1/\sqrt{2l+1}$ generic error bounds from cosmic variance. Agreements can be observed generally between the shapes and strength of the $\widetilde C\sups{(2,2)}_l$ and $\widetilde C_l$ curves, and between the two error estimation methods. The comparison in the relative errors of the two estimation methods can be found in \refig{C2r}.}{}{}
Even in a purely Gaussian universe, we should still expect a certain amount of power multipoles, which are naturally generated by the quantum fluctuations. This is confirmed in \refeq{pm-G-C}, which determines the power multipoles from $C_l$ solely. Given our current best estimate of the $C_l$ curve by the Planck observations \cite{Ade:2013zuv}, we can predict the power multipoles assuming our universe is Gaussian. The predicted power multipoles are shown in \refig{C2}, in terms of
\eq{\widetilde C\sups{(2,2)}_l\equiv \sqrt{l(l+1)C\sups{(2,2)}_l},}
where we have omitted the ``non-connected'' superscript. It is also compared with the curve for the best estimate of $C_l$, with the scaling
\eq{\widetilde C_l\equiv l(l+1)C_l.}
It can be seen from \refig{C2} that the power multipole signals should be strong and easily measurable. Its signal-to-noise ratio should be just half of that of $C_l$ from an isotropic Gaussian noise.

\ssecs{From cosmic variance}
It should be noted that power multipoles also suffer from cosmic variance, which limits observations and leaves large statistical errors, especially for low-$l$ power multipoles. Although $a_{lm}^{(2)}$ are not Gaussian variables, we can still expect the cosmic variance to have the same order of magnitude, i.e.\ $\sim 1/\sqrt{2l+1}$. Therefore, an inconsistency between theory and observation in the power dipole ($l=1$) should only be announced when the relative discrepancy reaches $\gtrsim {\cal O}(1)$. The high $l$ power multipoles, on the contrary, should be more sensitive indicators of any deviation from the standard paradigm of cosmology.

The CMB angular spectrum $C_l$ also suffers from cosmic variance, and its statistical error should also affect our expectation on the power multipoles. However the cosmic variances of $C_l$ and $C\sups{(2,2)}_l$ should not be regarded as independent, because in any specific universe the CMB power multipoles we see are fully determined by the CMB multipoles of the same universe.

\fig{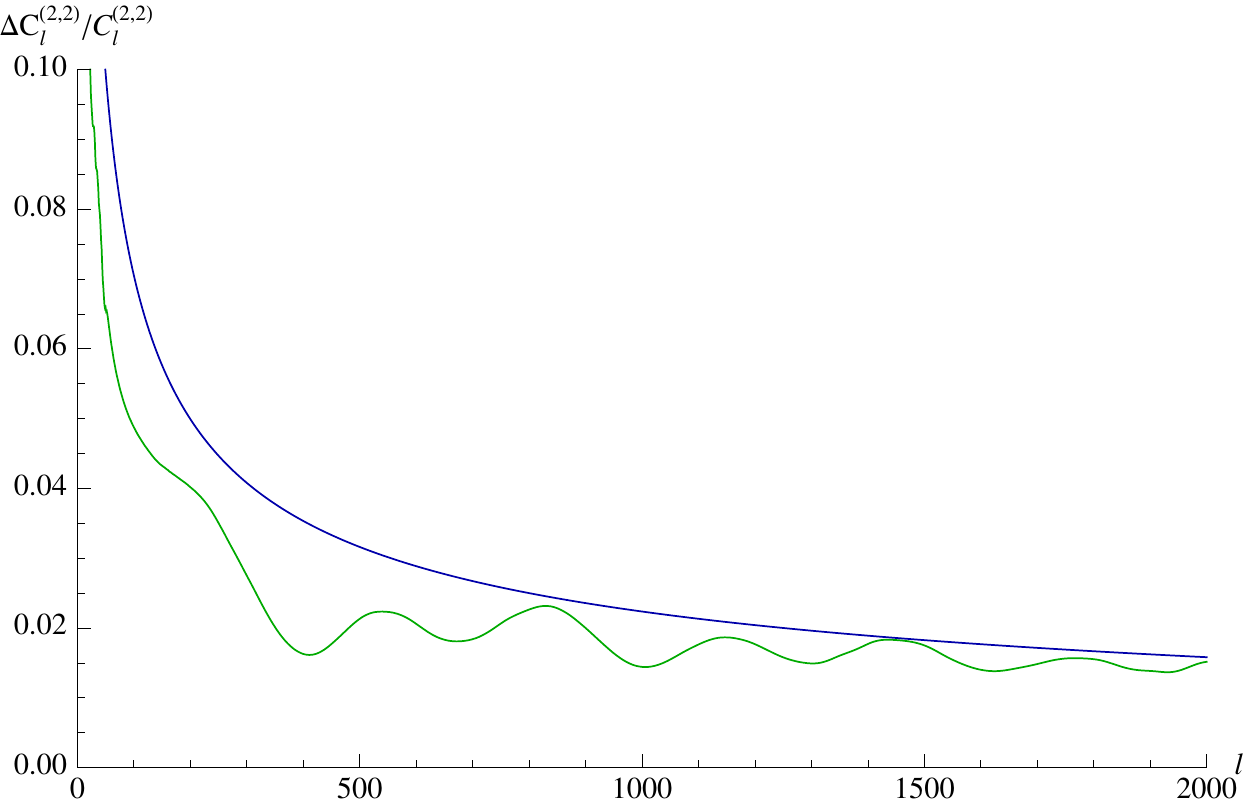}{The generic $1/\sqrt{2l+1}$ error estimation (blue) of cosmic variance is shown to agree well with the Monte Carlo simulation (green) at $1\sigma$.}{0.8}{}
We can even perform a simple numerical estimation by considering the following scenario.
\footnote{While analytical calculations of the cosmic variance effect is possible for a Gaussian universe, it would introduce additional complexity from the eight-point angular correlation functions of the CMB fluctuations. In this article, I simply choose Monte Carlo simulations as the numerical workaround.}
Suppose the CMB spectrum of the universe ensemble is $C_l$, which is impossible to know exactly. We can only know their best estimate by observing our universe, and apply these values to infer the power multipoles, or $C\sups{(2,2)}_l$. This induces errors between the the actual $C\sups{(2,2)}_l$ of the universe ensemble and our inferred one, which can be estimated with the standard deviation\footnote{
For simplicity, this paper neglects Bessel corrections in the large-$N$ limit.}
\eq{\Delta C\sups{(2,2)}_l\equiv\sqrt{\left\langle\Bigl(\left.C\sups{(2,2)}_l\right|\subs{inferred}-\left.C\sups{(2,2)}_l\right|\subs{true}\Bigr)^2\right\rangle}.}
This can be numerically calculated through Monte Carlo simulations, by taking the true $C_l$ values as we observe, in a naive but cost-efficient manner. By simulating $10^5$ universes, the error bars are obtained in \refig{C2}, or the relative errors in \refig{C2r}. The errors agree very well with the generic estimate $1/\sqrt{2l+1}$, suggesting the two cosmic variances are correlated and the $1/\sqrt{2l+1}$ works as a good estimation.

\ssecs{From non-Gaussian perturbations}
If the CMB power multipoles agree well with \refig{C2}, we then see no difference between our current universe and the simplest Gaussian universe. Only if we find a significant deviation, should we investigate the possible origins of the excessive power multipoles.

There are many possible sources of the power multipoles in reality, such as systematic or instrumental errors. Here we assume the observational groups have taken good care of them, so the excess would indicate a deviation from the standard paradigm of cosmology. According to \refeq{pm-pm}, the power multipoles are determined by the angular trispectrum of the CMB, so we can find a simple explanation of any excessive power multipoles from non-Gaussian temperature fluctuations.

In case of any primordial non-Gaussianity, in terms of $\tNL$ or $\gNL$, the angular trispectrum receives contribution in the connected part, while the non-connected part does not change at tree level. The connected angular trispectrum is calculated in \refsec{Angular trispectra from primordial non-Gaussianities}, which leads to the extra power multipoles by the amount

\eqa{C\sups{(2,2,c)}_l&=&\frac{25\tNL}{512\pi^2}\sum_{l_1l_2l_3l_4L}A^{l_1l_2}_{l_3l_4}(L)\sqrt{P_{l_1l_2L}P_{l_3l_4L}}\Biggl(\,\prod_{n=1}^4\sqrt{2l_n+1}\Biggr)\nonumber\\
&\times&\Biggl(\sqrt{P_{l_1l_2l}P_{l_3l_4l}}\delta_{Ll}+(-1)^{l+L}(2L+1)\Biggl(\sqrt{P_{l_1l_4l}P_{l_2l_3l}}\Sj{l_1}{l_2}{L}{l_3}{l_4}{l}+\underset{l_3\leftrightarrow l_4}{\mathrm{perm}}\Biggr)\Biggr)\nonumber\\
&+&\frac{3\gNL}{64\pi}\sum_{l_1l_2l_3l_4}\Biggl(\,\prod_{n=1}^42l_n+1\Biggr)P_{l_1l_2l}P_{l_3l_4l}B_{l_1l_2l_3l_4},}
where the $6j$ symbol is defined in \refssec{6j symbols}.

In this sense, the modulation measurements can also detect CMB non-Gaussianities through power multipoles. This will not reproduce all the possible degrees of freedom (${\cal O}(l\subs{max}^5)$ where $l\subs{max}\sim2500$ for Planck is the angular resolution) for the angular trispectra, but its ${\cal O}(l\subs{max})$ linear combinations can already provide additional information than simply the $\tNL$ or $\gNL$ directions. Measuring the shape of the $C\sups{(2,2)}_l$ curve allows for further inspections on the trispectra shapes, especially those which are (almost) orthogonal to $\tNL$ and $\gNL$, and are regarded as cancellations sometimes. It can also distinguish other possible deviations from the standard paradigm of cosmology, such as the break in the scale-invariance at very large scales \cite{Mazumdar:2013yta}, because the information of $P_\zeta(k)$ is contained in $A^{l_1l_2}_{l_3l_4}(L)$ and $B_{l_1l_2l_3l_4}$.

\secs{Comparison with CMB modulation models}
The dipole modulation model was first proposed in \cite{Eriksen:2007pc} to explain the observed power asymmetry, by adding a phenomenological dipole modulation in the form
\eq{\Delta T(\uv{n})=(1+A\,\uv{p}\cdot\uv{n})\Delta T\subs{iso}(\uv{n}),\label{eq-dm-Dt}}
where the dipole modulation with the strength $A$ along the direction $\uv p$, acts upon the isotropic Gaussian temperature fluctuations $\Delta T\subs{iso}(\uv n)$.

A generic modulation has also been discussed in \cite{Ade:2013nlj}, in the form
\eq{\Delta T(\uv{n})=(1+M(\uv{n}))\Delta T\subs{iso}(\uv{n}),\label{eq-dm-Dtg}}
so the temperature fluctuations are modulated by the generic function $M(\uv{n})$, which can contain any multipole. Therefore \refeq{dm-Dtg} is a complete parameterization of all the possible linear modulations in the CMB.

The newly introduced modulation parameters, $A$ or $M(\uv n)$, are regarded as free parameters which \emph{break statistical isotropy artificially}. Therefore we can measure the violation of statistical isotropy to detect the CMB modulations, through the Bipolar Spherical Harmonics (BipoSH) method \cite{Hajian:2003qq}.

On the contrary, the power multipoles \emph{assume statistical isotropy} in the universe ensemble, and tries to identify other possible sources of the CMB modulations, such as a small or mild non-Gaussianity, while maintaining statistical isotropy. It also relies on the more fundamental observables $\Delta T^2(\uv n)$ which, after assuming statistical isotropy, become solely determined by the primordial cosmology without the introduction of any ad-hoc parameter. It should also allow the reuse of some of the techniques for analyzing the CMB multipoles, such as galaxy masks.

Despite that the two methods address different sources of the CMB modulation, it is still possible to perform a floppy comparison between the acurracies of the two measurements, although a strict one under identical conditions would be impossible. Consider the generic form of modulation \refeq{dm-Dtg}. The CMB perturbation power becomes $\Delta T^2(\uv n)=(1+2M(\uv n))\Delta T\subs{iso}^2(\uv n)$ up to first order of $M(\uv n)$. By smoothening the CMB power map, we can approximately replace $\Delta T\subs{iso}^2(\uv n)$ with its expectation $\langle\Delta T\subs{iso}^2(\uv n)\rangle$. The modulation $M(\uv n)$ then produces the CMB power multipoles
\eq{a_{lm}^{(2)}\sim2M_{lm}\langle\Delta T\subs{iso}^2(\uv n)\rangle,}
where $M_{lm}\ll1$ is the spherical harmonic expansion of $M(\uv n)=\sum_{lm}M_{lm}Y_{lm}(\uv n)$. Because of statistical isotropy, we should have $\langle|M_{lm}|^2\rangle=M_l^2$ for any $m$, and thus the power multipoles from the modulation $M(\uv n)$ can be estimated as
\eq{C^{(2,2)}_l\sim4M_l^2\langle\Delta T\subs{iso}^2(\uv n)\rangle^2.}

\fig{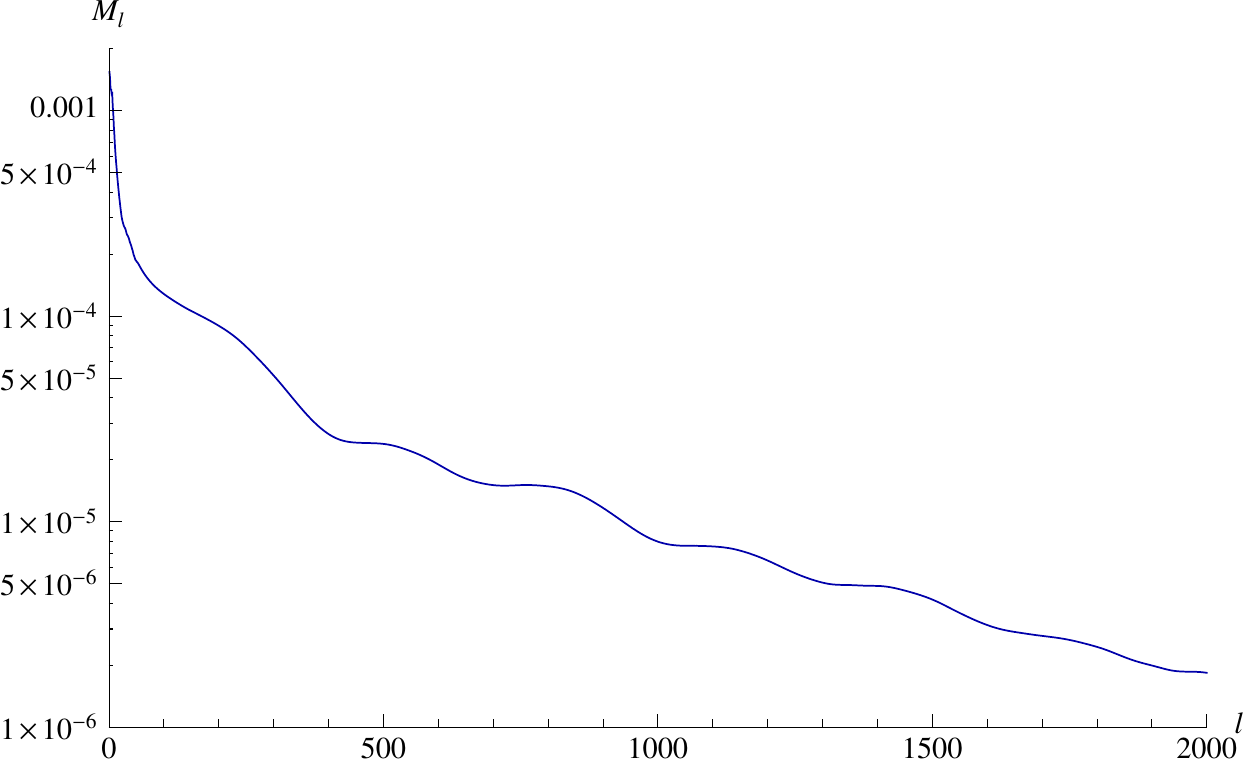}{The theoretical accuracy for the measurement of the CMB modulations $M_l$ through the CMB power multipoles. The blue curve shows the estimated statistical error of $M_l$ from cosmic variance at $1\sigma$ significance. Other sources of the measurement errors, such as noise contributions, are not considered in the figure.}{0.8}{}
The cosmic variance prevents accurate measurements on the power multipoles. According to \refssec{From cosmic variance}, we are only able to measure excessive power multipoles that is greater than the cosmic variance effect, to the amount $\gtrsim C_l\sups{(2,2,nc)}/\sqrt{2l+1}$. This corresponds to the theoretical accuracy
\eq{M_l\gtrsim\frac{\Bigl(C_l\sups{(2,2,nc)}\Bigr)^\frac{1}{2}}{2(2l+1)^\frac{1}{4}\langle\Delta T\subs{iso}^2(\uv n)\rangle}.}

We can even employ the more accurate results from Monte Carlo simulations, based on which the estimated accuracy for $M_l$ is predicted in \refig{C2e}. When noise is absent, the estimated accuracy tends to be higher for higher $l$, because the background signal and the cosmic variance effect are both weaker. This calls for the high $l$ modulation measurements to detect possible deviations from the standard paradigm of cosmology. For low $l$, the power multipole method also acquires a higher accuracy than the modulation model measurements using BipoSH, such as the accuracy for $M_1$ (or $A$) is $\sim0.0015$, which is much smaller than $0.02$ from \cite{Hanson:2009gu,Ade:2013nlj}. Because the non-Gaussianity measurements (e.g.\ $\tNL$) of the Planck observation rely on the detection of the CMB modulations  \cite{Ade:2013ydc}, a more accurate modulation measurement should also be able to further constrain the primordial non-Gaussianities.

It should be kept in mind that this section and \refig{C2e} are only order of magnitude estimations in theory, using the smoothing approximation. Actual measurements which include all the practical factors are still needed in reality to predict its accuracy and even measure the CMB power multipoles. Nevertheless, any isotropic Gaussian noise should not ruin the predictions because the signal-to-noise ratio of the CMB power map should be just half of that of the CMB map.

\secs{Conclusion}
In this paper, I have proposed the CMB power multipoles as a model-independent measurement of the CMB modulations. Assuming statistical isotropy, theoretical studies have shown that the CMB power multipoles are fully determined by the angular trispectra of the CMB, and thus should yield to \refig{C2} for our Gaussian universe with the cosmic variance taken into account. The CMB power multipoles then act as null tests of the standard paradigm of cosmology, although performing the actual tests with observational data is beyond the scope of this paper.

The standard paradigm is challenged when excessive power multipoles are observed to be greater than the cosmic variance effect. This would indicate a detection of CMB trispectra, whose shapes can be further determined through the measured power multipole curve, distinguishing between the sources such as $\tNL$, $\gNL$ and the violation of scale invariance at very large scales with a small non-Gaussianity.

On the other hand, if the observed power multipoles are consistent with the theoretical predictions, the CMB modulations, and hence the CMB non-Gaussianities, can be constrained. The power multipoles can place tighter bounds than the Planck observations, according to the brief theoretical estimate in \refsec{Comparison with CMB modulation models}, such as the dipole modulation can typically reach the accuracy $\sim 0.0015$. The tighter bounds should also introduce stronger consequential constraints for the primordial CMB trispectra.

The above conclusions are only reached in the ideal circumstance, i.e.\ watching the full sky without any noise or secondary CMB effect. In reality, they may affect the above predictions, although the sky masks are expected to apply similarly with the CMB multipole analysis, and any isotropic Gaussian noise are not expected to play any important role. Based on the promising accuracy of the modulation measurements and the consequential improved constraints on primordial CMB trispectra, it would be worthwhile to measure the CMB power multipoles from the observational perspective, and investigate it in greater detail in theory.

\acknowledgments{}
I would like to thank Ruth Durrer, Samuel Flender, Shaun Hotchkiss, Antony Lewis, and Anupam Mazumdar for helpful discussions and/or valuable comments on the draft.

\appendix
\secs{Angular trispectra from primordial non-Gaussianities}
The connected primordial trispectrum of the primordial curvature perturbation $\Phi$ is written in terms of $F\subs c(\vect k_1,\vect k_2,\vect k_3,\vect k_4)$, to express its connected four-point correlation function as
\eq{\langle\Phi(\vect k_1)\Phi(\vect k_2)\Phi(\vect k_3)\Phi(\vect k_4)\rangle\subs c=(2\pi)^3\delta^3(\vect k_1+\vect k_2+\vect k_3+\vect k_4)F\subs c(\vect k_1,\vect k_2,\vect k_3,\vect k_4).}
Two specific trispectrum shapes that receive most attention are parameterized in terms of $\tNL$ and $\gNL$, as \cite{Bartolo:2004if}
\eqa{F\subs c(\vect k_1,\vect k_2,\vect k_3,\vect k_4)&=&\frac{25}{16}\tNL\Biggl(P_\Phi(k_1)P_\Phi(k_{12})P_\Phi(k_3)+\underset{k_1,k_2,k_3,k_4}{23\ \mathrm{perms}}\Biggr)\nonumber\\&+&6\gNL\Biggl(P_\Phi(k_1)P_\Phi(k_2)P_\Phi(k_3)+\underset{k_4\leftrightarrow k_1,k_2,k_3}{3\ \mathrm{perms}}\Biggr),\label{eq-atri-Fc}}
where $\vect k_{12}\equiv\vect k_1+\vect k_2$.

Primordial non-Gaussianities coming from nonvanishing $\tNL$ or $\gNL$ will then contribute to the connected angular trispectrum, according to \refeq{bo-alm}, by the amount
\eqa{&&\langle a_{l_1m_1}a_{l_2m_2}a_{l_3m_3}a_{l_4m_4}\rangle\subs c\nonumber\\
&=&\frac{1}{2\pi^5}\int\delta^3(\vect k_1+\vect k_2+\vect k_3+\vect k_4)F\subs c(\vect k_1,\vect k_2,\vect k_3,\vect k_4)\prod_{n=1}^4(-i)^{l_n}Y^*_{l_nm_n}(\uv{k}_n)g_{l_n}(k_n)\dd^3\vect k_n.\label{eq-atri-ac}}
Contributions from $\tNL$ and $\gNL$ will be discussed separately.

\subsection{Angular trispectrum from $\tNL$}
Focusing on the first term in the permutation, we can rewrite the $\delta$ function as
\eq{\delta^3(\vect k_1+\vect k_2+\vect k_3+\vect k_4)=\int\dd^3\vect k_{12}\,\delta^3(\vect k_1+\vect k_2-\vect k_{12})\delta^3(\vect k_3+\vect k_4+\vect k_{12}).}
The two separate $\delta$ funtions can then be treated according to \cite{Wang:1999vf}, in terms of $j_l(x)$, the spherical Bessel functions, giving
\eqa{&&\delta^3(\vect k_1+\vect k_2+\vect k_3+\vect k_4)\nonumber\\
&=&64\int k_{12}^2y_1^2y_2^2\dd k_{12}\dd y_1\dd y_2\sum_{LM}(-1)^Mj_L(k_{12}y_1)j_L(k_{12}y_2)\nonumber\\
&\times&\sum_{{}^{\tilde l_1\tilde m_1\tilde l_2\tilde m_2}_{\tilde l_3\tilde m_3\tilde l_4\tilde m_4}}G_{\tilde l_1\,\ \tilde l_2\,\ L}^{\tilde m_1\tilde m_2M}G_{\tilde l_3\,\ \tilde l_4\ \,\ L}^{\tilde m_3\tilde m_4-M}\prod_{n=1}^4(-i)^{\tilde l_n}Y_{\tilde l_n\tilde m_n}(\uv{k}_n)j_{\tilde l_n}(k_ny_{(n)}),}
where $y_{(1)}=y_{(2)}\equiv y_1$ and $y_{(3)}=y_{(4)}\equiv y_2$.

For each of the permutations in \refeq{atri-Fc}, the $\delta$ function can be treated accordingly. After plugging the $\delta$ function expressions into \refeq{atri-ac}, we can find
\eqa{&&\langle a_{l_1m_1}a_{l_2m_2}a_{l_3m_3}a_{l_4m_4}\rangle\subs c\nonumber\\
&=&\frac{25}{8}\tNL\sum_{LM}(-1)^{M+l_1+l_2+l_3+l_4}\Biggl(A^{l_1l_2}_{l_3l_4}(L)G^{m_1m_2M}_{l_1\ \,l_2\ \,L}G^{m_3m_4-M}_{l_3\ \,l_4\ \ L}+\underset{l_2\leftrightarrow l_3,l_4}{2\mathrm{\ perms}}\Biggr),}
where
\cm{\eqa{A^{l_1l_2}_{l_3l_4}(L)&\equiv&\int\dd y_1\dd y_2\,y_1^2\alpha_{l_1}(y_1)\beta_{l_2}(y_1)\gamma_L(y_1,y_2)\beta_{l_3}(y_2)\alpha_{l_4}(y_2)y_2^2\nonumber\\
&+&3\mathrm{\ perms\ }(l_1\leftrightarrow l_2\times l_3\leftrightarrow l_4),}}
\eq{A^{l_1l_2}_{l_3l_4}(L)\equiv\int k^2\dd k\,P_\Phi(k)\gamma_{l_1l_2,L}(k)\gamma_{l_3l_4,L}(k),}
and
\eqa{\alpha_l(y)&\equiv&\frac{2}{\pi}\int k^2\dd k\,g_l(k)j_l(ky),\\
\beta_l(y)&\equiv&\frac{2}{\pi}\int k^2\dd k\,P_\Phi(k)g_l(k)j_l(ky),\\
\cm{\gamma_l(y_1,y_2)&\equiv&\frac{2}{\pi}\int k^2\dd k\,P_\Phi(k)j_l(ky_1)j_l(ky_2)}
\gamma_{l_1l_2,l}(k)&\equiv&\sqrt\frac{2}{\pi}\int\dd y\,y^2j_l(ky)\Biggl(\alpha_{l_1}(y)\beta_{l_2}(y)+\underset{l_1\leftrightarrow l_2}{\mathrm{perm}}\Biggr).}

\subsection{Angular trispectrum from $\gNL$}
Similarly, we can rewrite the $\delta$ function as
\eqa{&&\delta^3(\vect k_1+\vect k_2+\vect k_3+\vect k_4)\nonumber\\
&=&32\pi\int y^2\dd y\sum_{{}^{\tilde l_1\tilde m_1\tilde l_2\tilde m_2L}_{\tilde l_3\tilde m_3\tilde l_4\tilde m_4M}}(-1)^M
G_{\tilde l_1\ \tilde l_2\ \;L}^{\tilde m_1\tilde m_2M}G_{\tilde l_3\ \tilde l_4\ \ L}^{\tilde m_3\tilde m_4-M}\prod_{n=1}^4j_{\tilde l_n}(k_n y)Y_{\tilde l_n\tilde m_n}(\uv k_n).}
This gives
\eq{\langle a_{l_1m_1}a_{l_2m_2}a_{l_3m_3}a_{l_4m_4}\rangle\subs c=3\pi\gNL\sum_{LM}(-1)^MB_{l_1l_2l_3l_4}G^{m_1m_2M}_{l_1\ \,l_2\ \,L}G^{m_3m_4-M}_{l_3\ \,l_4\ \ L},}
where
\eq{B_{l_1l_2l_3l_4}\equiv\int y^2\dd y\,\alpha_{l_1}(y)\beta_{l_2}(y)\beta_{l_3}(y)\beta_{l_4}(y)+\underset{l_1\leftrightarrow l_2,l_3,l_4}{3\mathrm{\ perms}}.}

\secs{Special functions}
\ssecs{3j symbols}
The $3j$ symbol characterizes the coupling between angular momenta. Its detailed definition and properties can be found in \cite{Bartolo:2004if,Olver:2010:NHMF,NIST:DLMF}. Here we only list the related ones.
\begin{enumerate}
\item Triangle conditions

The $3j$ symbol $\Wj{1}{2}{3}$ is nonvanishing only if all of the following conditions are met
\begin{itemize}
\item $2l_1,2l_2,2l_3\in\mathbb{N}_0$.
\item $|l_1-l_2|\le l_3\le l_1+l_2$.
\item $m_i=-l_i,-l_i+1,\dots,l_i$, for $i=1,2,3$.
\item $m_1+m_2+m_3=0$.
\end{itemize}
In this paper, the spherical harmonic expansion enforces a stronger constraint $l_1,l_2,l_3\in\mathbb{N}_0$.
\item Symmetries
\eqa{\Wj{1}{2}{3}&=&\Wj{2}{3}{1}=\Wj{3}{1}{2},\\
(-1)^{l_1+l_2+l_3}\Wj{1}{2}{3}&=&\Wj{1}{3}{2}=\Wigner{l_1}{l_2}{l_3}{-m_1}{-m_2}{-m_3}.}
\item Orthogonalities
\eqa{\sum_{m_1m_2}(2l_3+1)\Wj{1}{2}{3}\Wigner{l_1}{l_2}{\tilde l_3}{m_1}{m_2}{\widetilde{m}_3}&=&\delta_{l_3\tilde l_3}\delta_{m_3\widetilde m_3},\\
\sum_{l_3m_3}(2l_3+1)\Wj{1}{2}{3}\Wigner{l_1}{l_2}{l_3}{\widetilde m_1}{\widetilde m_2}{m_3}&=&\delta_{m_1\widetilde m_1}\delta_{m_2\widetilde m_2}.}
\item Other relations
\eqa{\Ga{1}{2}{3}&=&\sqrt\frac{(2l_1+1)(2l_2+1)(2l_3+1)}{4\pi}\Wjz{1}{2}{3}\Wj{1}{2}{3},\\
\Wjz{1}{2}{3}^2&=&\frac{1}{2}P_{l_1l_2l_3}\nonumber\\
&=&\left\{\begin{array}{l@{\hspace{0.3in}}l}
0,&L\mathrm{\ odd},\\
\displaystyle\frac{(L-2l_1)!(L-2l_2)!(L-2l_3)!(L/2)!^2}{(L+1)!(L/2-l_1)!^2(L/2-l_2)!^2(L/2-l_3)!^2},&L\mathrm{\ even},\end{array}
\right.\hspace{0.3in}}
where $L\equiv l_1+l_2+l_3$ in the above expression.
\end{enumerate}

\ssecs{6j symbols}
The $6j$ symbols can be defined through $3j$ symbols as
\eqa{\left\{\begin{array}{ccc}L_1&L_2&L_3\\l_1&l_2&l_3\end{array}\right\}&\equiv&\sum_{M_im_j}(-1)^{l_1+l_2+l_3+m_1+m_2+m_3}\Wigner{L_1}{L_2}{L_3}{M_1}{M_2}{M_3}\nonumber\\
&&\times\Wigner{L_1}{l_2}{l_3}{M_1}{m_2}{-m_3}\Wigner{l_1}{L_2}{l_3}{-m_1}{M_2}{m_3}\Wigner{l_1}{l_2}{L_3}{m_1}{-m_2}{M_3},}
where $i,j=1,2,3$. More properties can also be found in \cite{Bartolo:2004if,Olver:2010:NHMF,NIST:DLMF}.

\bibliographystyle{jcap}
\bibliography{Main}
\end{document}